\documentclass{nature}

\usepackage{amssymb}
\usepackage{graphicx}
\usepackage{times}
\usepackage{epsfig}
\usepackage{amsmath}
\usepackage{subfigure}
\usepackage{amssymb}
\usepackage{epstopdf}
\usepackage[font=scriptsize]{caption}
\usepackage{float}
\usepackage[]{algorithm2e}

\title{Automated Body Structure Extraction from Arbitrary 3D Mesh}

\author{Yong Khoo, Sang Chung}

\begin{document}

\maketitle

\begin{abstract}
This paper presents an automated method for 3D character skeleton extraction that can be applied for generic 3D shapes. Our work is motivated by the skeleton-based prior work on automatic rigging focused on skeleton extraction and can automatically aligns the extracted structure to fit the 3D shape of the given 3D mesh. The body mesh can be subsequently skinned based on the extracted skeleton and thus enables rigging process. In the experiment,  we apply public dataset to drive the estimated skeleton from different body shapes, as well as the real data obtained from 3D scanning systems. Satisfactory results are obtained compared to the existing approaches. 

\end{abstract}

\section*{Introduction}
Nowadays, most of 3D animation tools enable people to create a skeleton of a 3D model manually. Although this technique provides great help to 3D animation, especially some strange 3D objects, it is not friendly enough to novices and time-consuming when a large number of 3D characters are in need.  Apart from creating a new easy-learn and convenient system for 3D animation, automatic generation of skeletons for 3D object is also a choice. In the first paper, Wade and Parent\cite{iref1} introduce an algorithm for automated construction of skeleton from mesh data. In the second paper, Baran and Popovic\cite{iref2} create a system, Pinocchio, that construct a skeleton and embed it into the characters for the use of animation.

\section*{Related Works}
Comparing to the previous review paper, I focus on 3D reconstruction. They demonstrate different method on building a 3D model. In this paper, it is review on the study how to animate the 3D object. Distance map, mentioned in both papers, is an important tool to the procedure of discretization of the 3D model. Early in 1998, Gagvani et al. \cite{iref3}  use distance map, called in distance transform in this paper, in animating 3D model with skeleton tree. For more details on how distance map interacts with computer graphics, Frisken et al.'s work\cite{iref4}  has a clear illustration on them.\\
According to Baran and Popovic, more and more researchers, taking the novice users into account, created an animation system that can used by users who have little knowledge on animation. In 2005, Shen et al. \cite{iref5}  present a system enable users to make a short 3D animation by controlling the 3D object to perform different move by fingers. More, they also create other similar systems\cite{iref6}, \cite{iref6b} for 3D animation. The second system record different points that indicating a variety of a 2D character's motions. By changing the points, the screen shows different motions of the character as different frames.\\
In most of works on this field, researches use skeleton extraction to find the skeletal graph. 
In 2003, Liu et al. \cite{iref7}  construct a repulsive force filed in a given model to generate the skeletal graph of the model. Same year, Katz and Tal\cite{iref8}  extract the skeleton of an object by segmenting it.\\
Skinning is one of the big problem for 3D animation. A good skinning work improves the quality of 3D animation and can facilitate many VR based applications \cite{iref8a}. In 2004, Yu et al.\cite{iref9}  employ the Poisson equation to manipulate 3D models. In 2005, Lipman et al.\cite{iref10}  introduce an approach on editing mesh based on object surface.

\section*{Method 1}
This section discusses the method of the first paper. The algorithm includes four steps: voxelization and distance map construction, medial surface extraction, path tree generation, and control skeleton construction.\\
\subsection{Voxelization and distance map construction}
In the first step, the researchers put a 3D model in to a box that can include all the model inside. The box is divided into small cubes in same size. The 3D distance map of the model is constructed. Figure 1 shows how to construct a 2D Euclidean distance map (EMD.\begin{figure*}[h]
\centerline{\epsfig{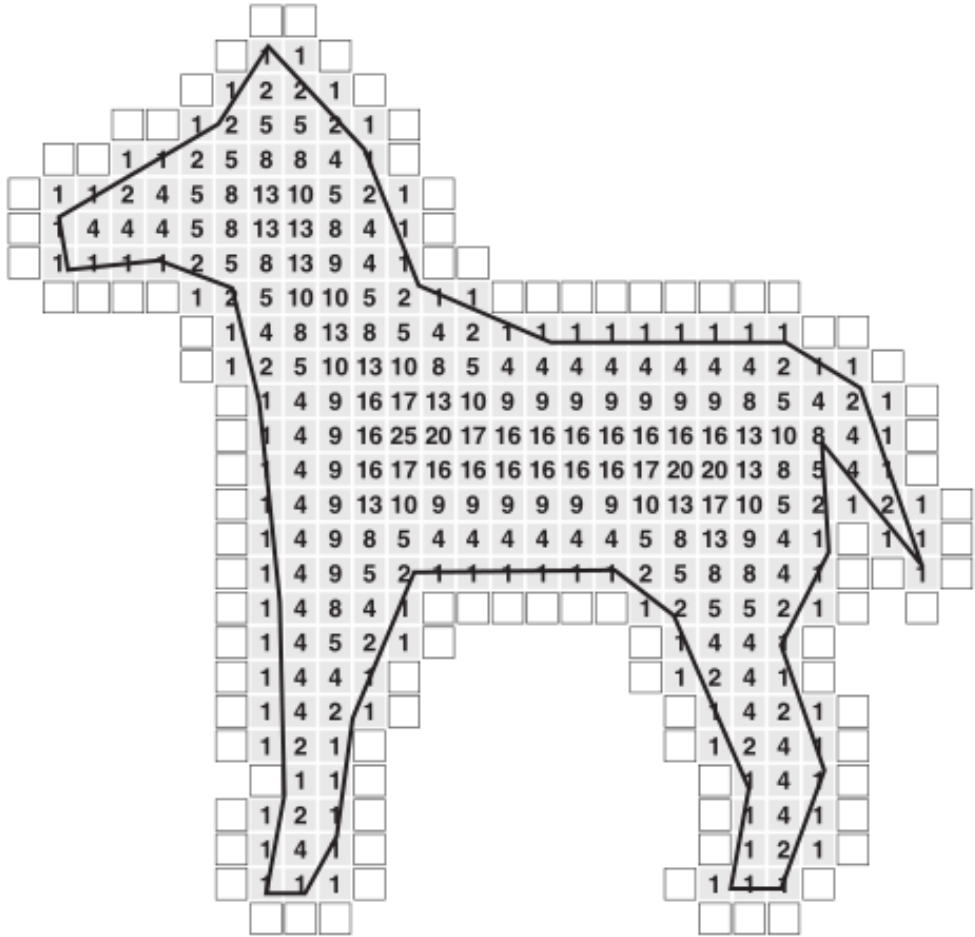}}
\caption{The 2D Euclidean distance map for a discretized horse polygon.}
\end{figure*}The number of a square means the value of the shortest distance from the square to the boundary.
\subsection{Medial surface extraction}
The EDM calculated in previous step is used to compute the discrete medial surface (DMS) \cite{iref6c}. Extraction algorithm is applied in this step. Daniellsson's\cite{iref11}, and Ge and Fitzpatrick's\cite{iref12} works are suggested to read for the detail on the extraction algorithm. The DMS result is show in Figure 2a.
\subsection{Path tree generation}
In this step, the heart of the DMS is found, which is a globally central voxel in the object, shown in Figure 2a. The researchers use a breadth-first search on the DMS to find extreme points, which are the local maximum points.\\
$$W_p=\sum_{v_i \in P}\frac{1}{d_i^3}$$\\
The procedure of path tree generation is show in Figure 2.\begin{figure*}[h]
\centerline{\epsfig{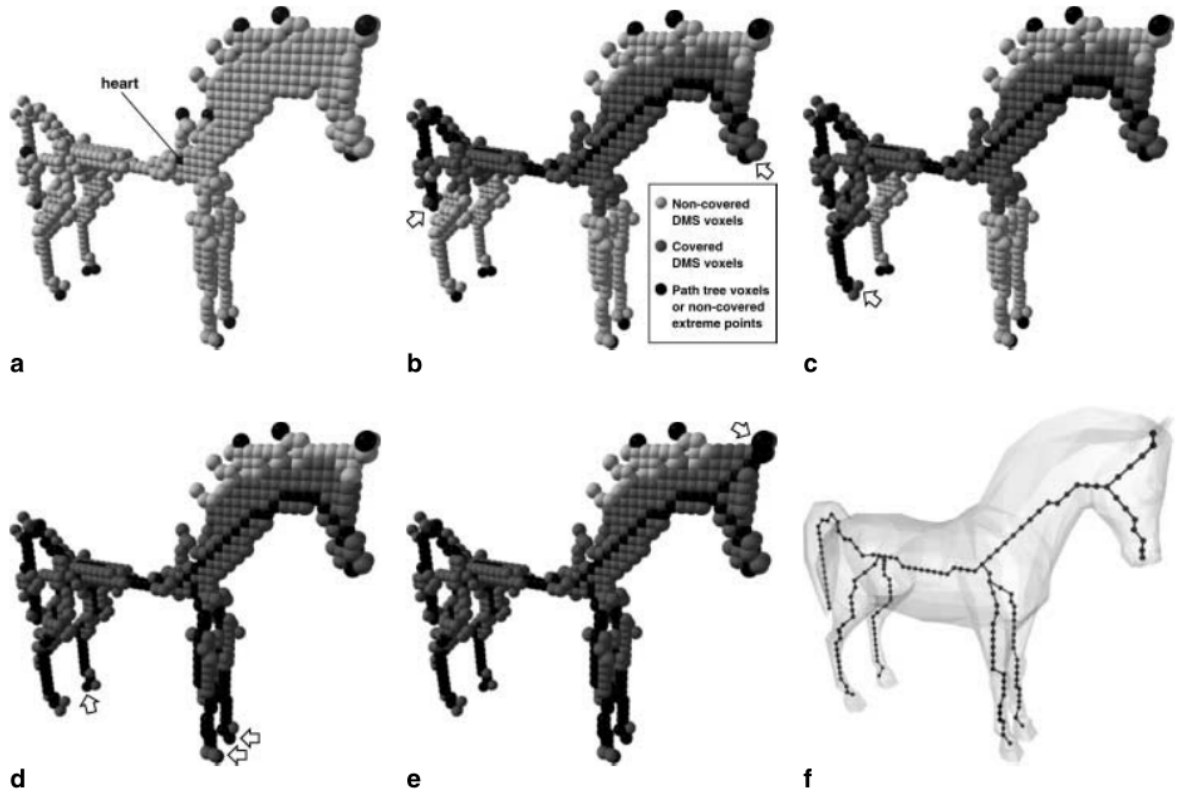}}
\caption{From a to f, they show the process of generating the path tree. The black points are path tree voxels or non-covered extreme points, the dark points are covered DMS voxels, and the grey points are non-covered DMS voxels.}
\end{figure*} The researchers, using the weighted measure function below, create a centralized path in the model. 
The path connects the farthest point to the heart through the voxels. The dark point region in the figure is made of covered DMS voxels. The covered DMS voxels are the voxels that in the overlap sphere whose center is on the path and whose radius equals to the shortest distance between the center and model's surface. Not all the extreme points are used to generate the path tree. Some of them are unqualified because the shortest path from it to a cover DMS voxel is less than the threshold. Then, researchers convert the jagged the path tree into the smooth one. The conversion is displayed in Figure 3 and Figure 4.\begin{figure*}[h]
\centerline{\epsfig{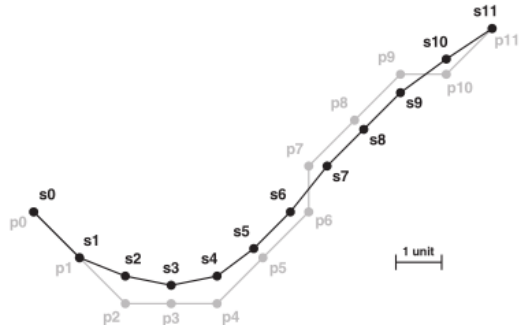}}
\caption{A smooth path-tree chain}
\end{figure*}\begin{figure*}[h]
\centerline{\epsfig{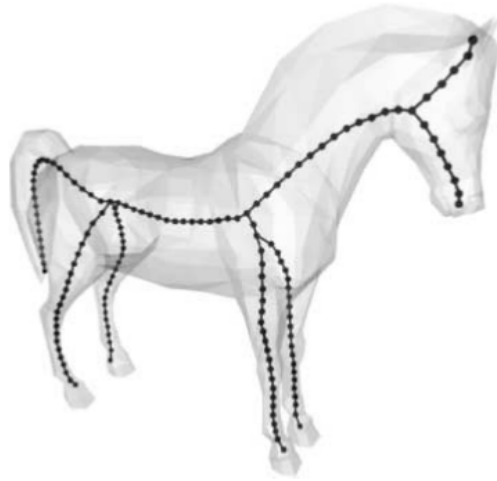}}
\caption{A path tree of the horse}
\end{figure*}
\subsection{Control skeleton construction}
The researchers using the path tree to construct the skeleton. Figure 5 illustrates the method on creating the skeletal graph. Figure 6 shows the horse skeleton constructed from the path tree.\begin{figure*}[h]
\centerline{\epsfig{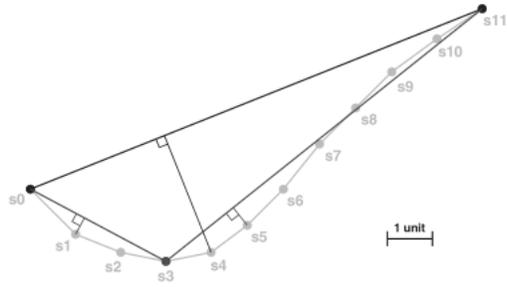}}
\caption{Splitting of a skeletal graph edge.}
\end{figure*}\begin{figure*}[h]
\centerline{\epsfig{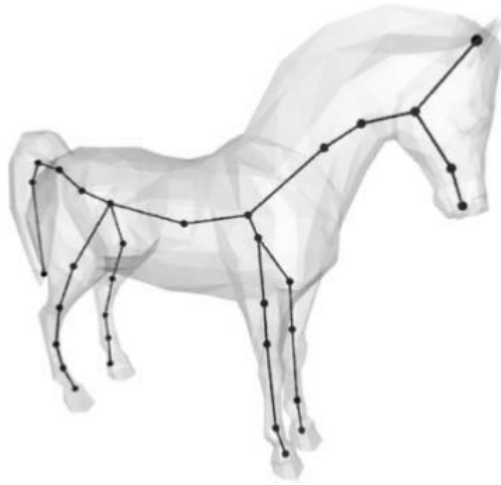}}
\caption{A skeleton of the horse}
\end{figure*} The basic idea is that an edge is split into two new edges repeatedly. There is an edge between two initial end points. The researchers use the points on the tree path that connects two initial end points as split points. Each split point can generate two edges with the end points. The split point has the smallest maximum error is chosen to be the new end point for the further splitting. The splitting stops when the number of segments parameters, depends on user's setting, is reached or the heap memory for dynamic use is empty. The new edges set is constructed as a skeletal graph. Every edge of the graph is a control segments of it. Each joint among the edge has three rotational degrees of freedom to change the certain part of the model. Finally, the problem is how to let the model move rationally under the control of the skeleton. Each chain in the graph manages its correspondent segments of the 3D model, which are based on the covered DMS voxels and the study on how covered DMS voxels influence the uncovered DMS voxels.

\section*{Method 2}
The method of the second paper is reviewed in this section. It contains two stages: skeleton embedding and skin attachment.\\
\subsection{Skeleton embedding}
The first step in this stage is discretization. The system builds a graph to represent a 3D object. Figure 7 shows the steps of creating the graph.\begin{figure*}[h]
\centerline{\epsfig{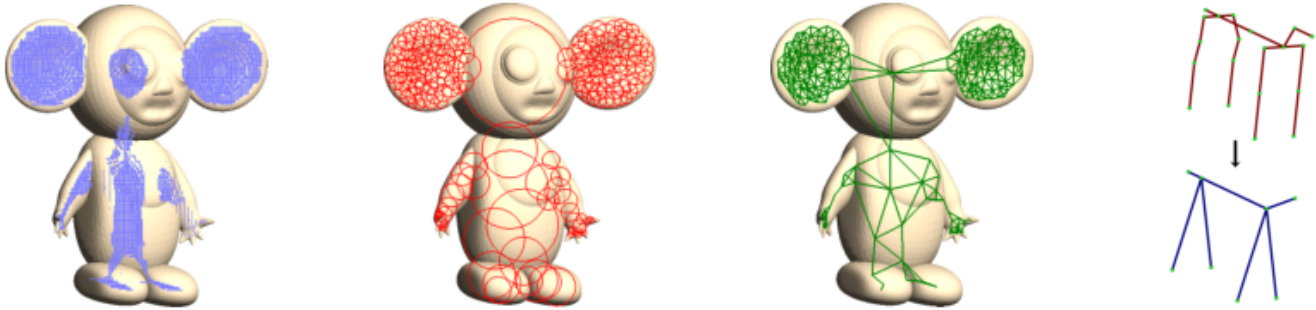}}
\caption{From left to righ, Figure 7a-7d, approximate medial surface, packed spheres, constructed graph, and The original and reduced skeleton.}
\end{figure*} The adaptive distance field is used in the system to build approximate medial surface. A graph point in the model that are farthest from the surface is selected. The point becomes the center of a sphere that fills the 3D object as much as possible and does not go beyond the surface. New graph points containing the same character are picked to build new spheres. More, each spheres do not contain the center of the others. Figure 7b show the object represented by packed spheres. Some pairs of centers are connected to construct the graph, showed in Figure 7c.\\
The graph is reduced and merged to construct the skeleton of the object. The graph is set as $G=(V,E)$. The reduced skeleton is represented by an r-tuple $\mathbf{v}=(v_1,...,v_r)$ of vertices in V, where $r$ means the joints in the reduced skeleton.The sample reduced skeleton, which has 7 joints, is showed in Figure 7d.\\
The discrete penalty function is proposed to ensure the good quality of the embedding by MRF based approach \cite{iref6d}. The researchers produce a penalty functions $f(\mathbf{v})=\sum^k_{i=1}\gamma_i b_i(\mathbf{v})$ to penalize the improper factors on the skeleton. The feature vector $\mathbf{b}$ is divided in to $\mathbf{p}$ and $\mathbf{q}$ where $\mathbf{p}$ representes the k-dimensional feature vectors of good embeddings and $\mathbf{q}$ does that of bad ones. Then, the researchers need to find $\Gamma$ that can maximum the margin between the best "bad" embedding and the best "good" embedding. The optimization margin is defined below. $$min^n_{i=1} \Gamma^T \mathbf{q}_i-min^m_{i=1} \Gamma^T \mathbf{p}_i           (||\Gamma||=1)$$
Figure 8 illustrate the optimization margin.
\begin{figure*}[h]
\centerline{\epsfig{figure=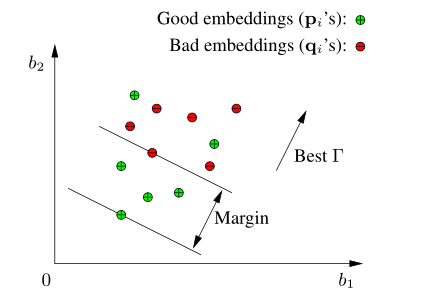,width=8cm}}
\caption{Figure 8}
\end{figure*}
The machine learning method is used in the optimization to find a stable $\Gamma$.\\
The next not the last step in this stage is discrete embedding. The whole embedding start from small partial embeddings. It sounds like jigsaw puzzle. Different partial embeddings have different priority.The best optimal part is embedded first.
Finally, the embedding is refined for better skeleton. Figure 9 show the embedding refinement.
\begin{figure*}[h]
\centerline{\epsfig{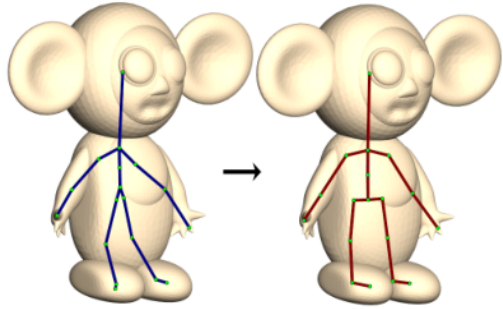}}
\caption{The blue skeleton change to the red one after embedding refinement}
\end{figure*}
\subsection{Skin attachment}
Only embedding skeleton into the object cannot present animation. Skin attachment is required to telling how the object deforms based on the skeleton. The standard linear blend skinning method is applied to the system. The function is shown here, $\sum_i w^i_j\mathbf{T}^i(\mathbf{v}_j)$. $\mathbf{v}_j$ means the position of vertex $j$, $\mathbf{T}^i$ is the transformation of the $i^{th}$ bone, and $w^i_j$ is the weight of the $i^{th}$ bone for vertex $j$. The problem here is to get bone weights $w^i$, which is the relationship between each bone and each vertex when deformation happens. The researchers use heat equilibrium method to solve the problem.

\section*{Result}
This section shows part of result in both paper.
\subsection{Result in the 1st paper}
The examples of result are showed in the Figure 5, horse, octopus and jellyfish.
\begin{figure*}[h]
\centerline{\epsfig{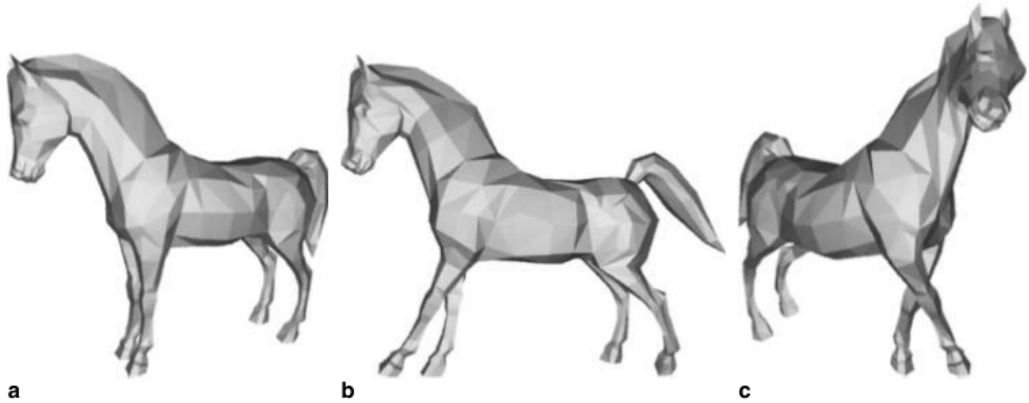}}
\caption{a. The default pose of the horse. The other are the horse in two random poses}
\end{figure*}
\begin{figure*}[h]
\centerline{\epsfig{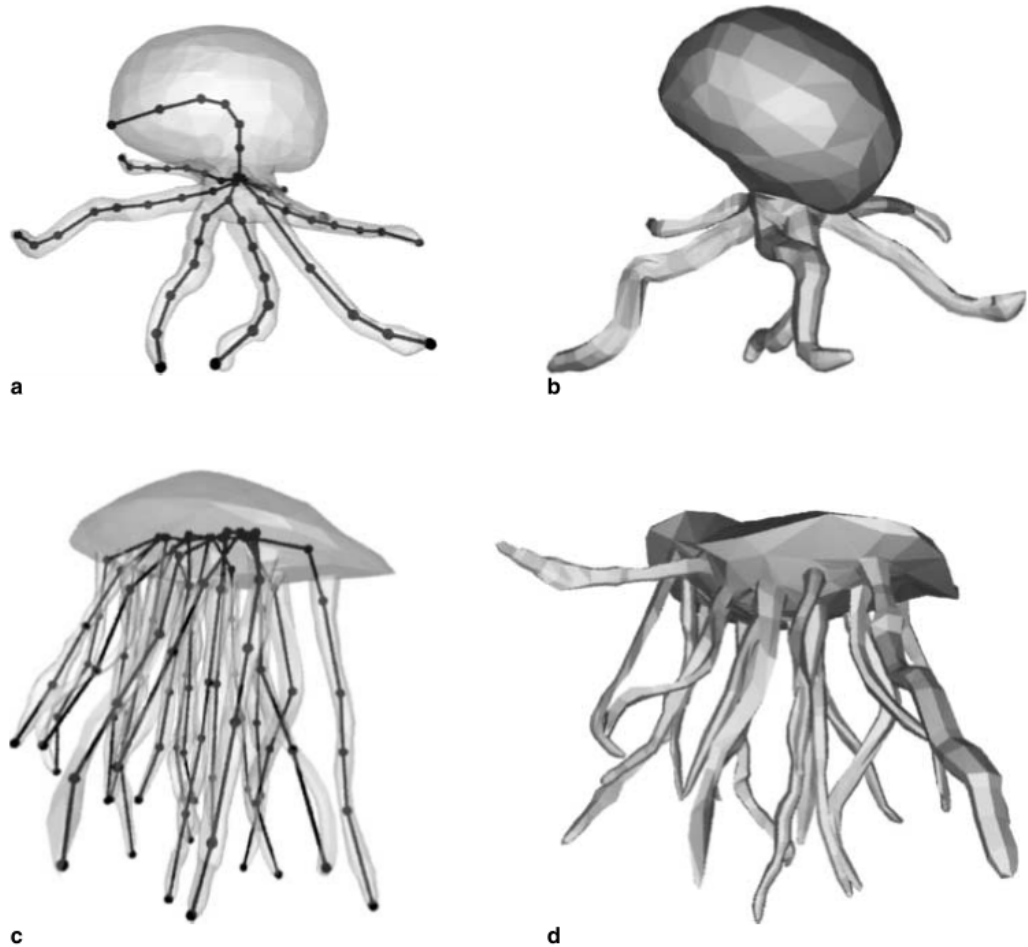}}
\caption{a. an octopus skeleton; b. an octopus pose; c. a jellyfish skeleton; d. a jellyfish pose.}
\end{figure*}

\subsection{Result in the 2nd paper}
Figure 6 below shows the test results of skeleton embedding. The 13th model of the test result suggest the limitation of the system.
\begin{figure*}[h]
\centerline{\epsfig{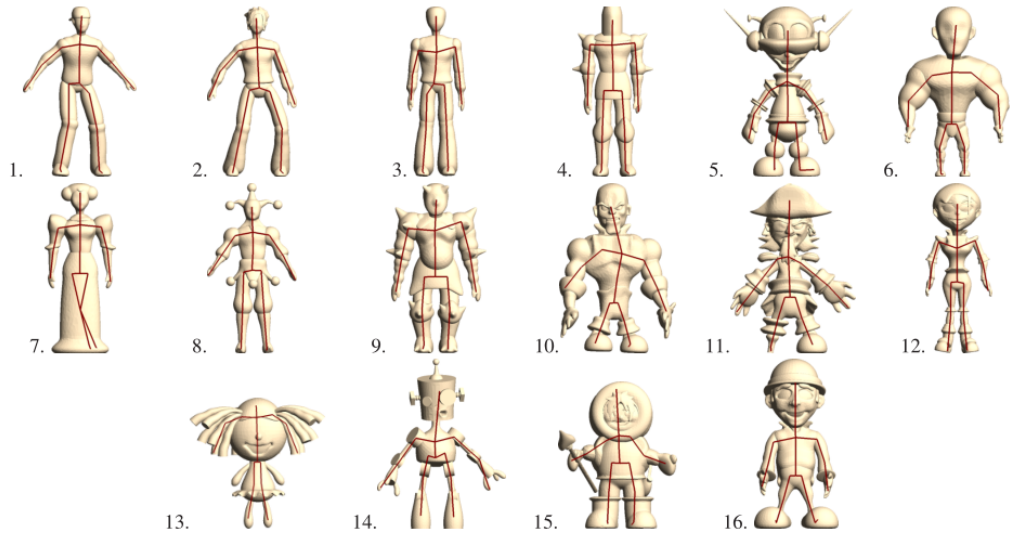}}
\caption{Test results of 16 models. The 13th model is an undesirable result produced by the system}
\end{figure*}

\end{document}